\documentclass[12pt,epsf]{article}
\usepackage{graphicx}
\begin{document}


\def\a{\alpha}
\def\b{\beta}
\def\c{\varepsilon}
\def\d{\delta}
\def\e{\epsilon}
\def\f{\phi}
\def\g{\gamma}
\def\h{\theta}
\def\k{\kappa}
\def\l{\lambda}
\def\m{\mu}
\def\n{\nu}
\def\p{\psi}
\def\q{\partial}
\def\r{\rho}
\def\s{\sigma}
\def\t{\tau}
\def\u{\upsilon}
\def\v{\varphi}
\def\w{\omega}
\def\x{\xi}
\def\y{\eta}
\def\z{\zeta}
\def\D{\Delta}
\def\G{\Gamma}
\def\H{\Theta}
\def\L{\Lambda}
\def\F{\Phi}
\def\P{\Psi}
\def\S{\Sigma}

\def\o{\over}
\def\beq{\begin{eqnarray}}
\def\eeq{\end{eqnarray}}
\newcommand{\gsim}{ \mathop{}_{\textstyle \sim}^{\textstyle >} }
\newcommand{\lsim}{ \mathop{}_{\textstyle \sim}^{\textstyle <} }
\newcommand{\vev}[1]{ \left\langle {#1} \right\rangle }
\newcommand{\bra}[1]{ \langle {#1} | }
\newcommand{\ket}[1]{ | {#1} \rangle }
\newcommand{\EV}{ {\rm eV} }
\newcommand{\KEV}{ {\rm keV} }
\newcommand{\MEV}{ {\rm MeV} }
\newcommand{\GEV}{ {\rm GeV} }
\newcommand{\TEV}{ {\rm TeV} }
\def\diag{\mathop{\rm diag}\nolimits}
\def\Spin{\mathop{\rm Spin}}
\def\SO{\mathop{\rm SO}}
\def\O{\mathop{\rm O}}
\def\SU{\mathop{\rm SU}}
\def\U{\mathop{\rm U}}
\def\Sp{\mathop{\rm Sp}}
\def\SL{\mathop{\rm SL}}
\def\tr{\mathop{\rm tr}}

\def\IJMP{Int.~J.~Mod.~Phys. }
\def\MPL{Mod.~Phys.~Lett. }
\def\NP{Nucl.~Phys. }
\def\PL{Phys.~Lett. }
\def\PR{Phys.~Rev. }
\def\PRL{Phys.~Rev.~Lett. }
\def\PTP{Prog.~Theor.~Phys. }
\def\ZP{Z.~Phys. }


\baselineskip 0.7cm

\begin{titlepage}

\begin{flushright}
UT-03-34
\end{flushright}

\vskip 1.35cm
\begin{center}
{\large \bf
Upper Bound On Gluino Mass From Thermal Leptogenesis 
}
\vskip 1.2cm
M.~Fujii${}^{1}$, M.~Ibe${}^{1}$ and T.~Yanagida${}^{1,2}$
\vskip 0.4cm

${}^1${\it Department of Physics, University of Tokyo,\\
     Tokyo 113-0033, Japan}

${}^2${\it Research Center for the Early Universe, University of Tokyo,\\
     Tokyo 113-0033, Japan}

\vskip 1.5cm

\abstract{
Thermal leptogenesis requires the reheating temperature $T_R \gsim 3\times 10^{9}$~GeV,   
which contradicts a recently obtained constraint on the reheating
temperature, $T_R \lsim 10^6$~GeV, for the gravitino mass of
 $100$~GeV$-10$~TeV.  
This stringent constraint comes from the fact that the hadronic decays
 of  gravitinos destroy very efficiently light elements produced by the
 Big-Bang nucleosynthesis.  
However, it is not applicable if the gravitino is the lightest
supersymmetric particle (LSP). 
We show that this solution to the  gravitino problem works for the case
 where the next LSP is a scalar charged lepton or a scalar neutrino.   
We point out that there is an upper bound on the gluino mass as 
$m_{\rm gluino} \lsim 1.8$~TeV so that the energy density of gravitino  
does not exceed the observed dark matter density 
$\Omega_{\rm DM}h^2\simeq 0.11$.} 
\end{center}
\end{titlepage}

\setcounter{page}{2}

\section{Introduction}

In the light of experimental data of neutrino oscillations the
leptogenesis \cite{Fukugita:1986hr} is the most interesting and fruitful
mechanism for explaining the baryon-number asymmetry in the universe. 
A detailed analysis on the thermal leptogenesis \cite{BBP} requires 
the reheating temperature $T_R \gsim 3\times 10^{9}$~GeV, which, however, 
leads to overproduction of unstable gravitinos \cite{Holtmann:1998gd}. 
Namely, decays of gravitinos produced after inflation destroy the
success of Big Bang nucleosynthesis (BBN) \cite{Weinberg:zq,Kawasaki:2000qr}. 
This problem is not solved even if one raises the gravitino mass 
$m_{3/2}$ up to 30~TeV \cite{Kawasaki}.
Thus, the thermal leptogenesis seems to have a problem with the
gravity mediation model of supersymmetry (SUSY) breaking. 
It is, however, pointed out in \cite{gravitinoLSP} that this gravitino
problem may be solved if the gravitino is the lightest SUSY particle
(LSP).\footnote{
The gauge mediation model provides a solution to the gravitino problem 
\cite{Fujii:2002fv}. 
In this letter we consider only the gravity mediation model
of SUSY breaking.
} 

We show, in this letter, that there is an upper bound on
gluino mass, $m_{\rm gluino} \lsim 1.8$~TeV, for the above solution to
work. 
This is because the heavier gluino produces more abundantly the
gravitino after the inflation and the mass density of the produced
gravitino LSP, $\Omega_{3/2}h^2$, exceeds the observed energy density of
dark matter, $\Omega_{\rm DM}h^2 \simeq 0.11$ \cite{DM}.  
The above bound on the gluino mass will be tested in the next generation
accelerator experiments  such as LHC.

We show that nonthermal gravitino production by decay of the next LSP
(NLSP) plays a crucial role of determining the upper bound on gluino
mass.  
We find that a consistent NLSP is a scalar charged  lepton or a scalar
neutrino.  
Its mass is stringently constrained as $m_{3/2} < m_{\rm NLSP} < m_{\rm
gluino}$, where the gravitino mass is in the region of $m_{3/2} \simeq
10-800$~GeV. 
Other candidates for the NLSP, that is, wino, Higgsino, scalar quark
(squark) and gluino, are excluded by a new strong constraint from the
BBN \cite{Kohri}, except for light gluino of mass $m_{\rm gluino} \lsim
70$~GeV. 
However, such a light gluino seems to be excluded by the present
accelerator experiments \cite{Baer:1998pg,Mafi:1999dg}. 

\section{Cosmological constraints and candidates for NLSP}

As we have mentioned in the introduction, even if the gravitino is the
stable LSP, it does not totally solve the cosmological gravitino problem
in the thermal leptogenesis. 
Namely, we have to constrain its relic abundance consistent with the
recent WMAP result \cite{DM}, $\Omega_{3/2}h^2\leq \Omega_{\rm DM}h^2 =
0.1126^{+0.0161}_{-0.0181}$. 
The production rate of the gravitino in the thermal bath depends on the
reheating temperature $T_R$ and the mass-squared ratio $m^2_{\rm
gluino}/m^2_{3/2}$. 
The resultant relic density of the  gravitino is
calculated as~\cite{Moroi,Buchmuller}   
\begin{equation}
\Omega_{3/2}^{\rm th}h^2\simeq 0.44\times \alpha_3(T_R)
\left(\frac{T_{R}}{10^{10}\GEV}\right)
\left(1+\frac{1}{3}\left(\frac{\alpha_3(T_{R})}{\alpha_3(\mu)}\right)^2
\left(\frac{m_{\rm gluino}}{m_{3/2}}\right)^2\right)
\left(\frac{m_{3/2}}{100\GEV}\right)
\label{eq:thermal_relic}\;,
\end{equation}
where $\alpha_3(\m)$ is a gauge coupling constant of SU$(3)_c$ at the
scale $\mu\simeq 1$~TeV and the
gluino mass $m_{\rm gluino}$ is the one given at the weak
scale.\footnote{    
Other SUSY particles such as wino and bino contribute to the gravitino
production. Thus, the result Eq.~(\ref{eq:thermal_relic}) is regarded as
the theoretical lower bound on the gravitino abundance.
}
The above WMAP constraint on $\Omega_{3/2}h^2\leq \Omega_{\rm DM}h^2$ 
gives an upper bound on the gluino mass  at a given reheating
temperature $T_{R}$ and a given gravitino mass $m_{3/2}$. 

Another important constraint comes from late-time decay of the NLSP into
a gravitino and its superpartner. The decay width of the NLSP is
approximately given by \cite{Moroi, Buchmuller} 
\begin{equation}
\Gamma_{\rm NLSP}\simeq \frac{1}{48 \pi}
\frac{m_{\rm NLSP}^5}{m_{3/2}^2 M_{*}^2}\;,
\label{eq:NLSP_decaywidth} 
\end{equation}
where $m_{\rm NLSP}$ is the mass of the NLSP and $M_{*}=2.4\times
10^{18}\GEV$ the reduced Planck scale. In terms of the lifetime,
it is written as
\begin{equation}
\tau_{\rm NLSP}\simeq 2.4\times 10^6 \;{\rm sec} 
\left(\frac{m_{3/2}}{100\GEV}\right)^2
\left(\frac{300\GEV}{m_{\rm NLSP}}\right)^5\;,
\label{eq:NLSP_lifetime}
\end{equation}
and then the NLSP can decay during or even after the BBN releasing a
large amount of energy.  
Therefore, we have to consider seriously effects of its decay on the BBN
to confirm the validity of a given model.

In order not to spoil the success of the BBN, we have to satisfy two
constraints on the abundance of the NLSP before its decay.
One of them comes from the hadronic energy release associated with the
NLSP decay. 
 Recently, a detailed analysis on the hadronic effects has been carried
 out for a wide range of the NLSP lifetime \cite{Kohri}.  
According to this research, even if we take a very conservative bound, 
the NLSP abundance is constrained as
\begin{equation}
B_{h}\times m_{\rm NLSP}Y_{\rm NLSP}\lsim 10^{-13}\GEV\quad{\mbox{for}}\quad
\tau_{\rm NLSP}
\gsim 10^{3}\;{\rm sec}\;.
\label{eq:hadron_constraint}
\end{equation}
Here, $B_{h}$ is the branching ratio of the NLSP decay into the hadroninc
components.  $Y_{\rm NLSP}$ is defined as the yield of the
NLSP before its decay, and is given by $Y_{\rm NLSP}\equiv n_{\rm
NLSP}/n_\gamma$, where $n_\gamma$ and $n_{\rm NLSP}$ are the densities
of the photon and the NLSP, respectively. 
Furthermore, if we take the constraint from $^{6}{\rm Li}/^{7}{\rm Li}$
into account, this constraint becomes much stronger as
\begin{equation}
B_{h}\times m_{\rm NLSP}Y_{\rm NLSP}\lsim 10^{-(15{\mbox{--}16})}\GEV\quad
{\mbox{for}}\quad 10^{3}\;{\rm sec}\lsim \tau_{\rm NLSP}\lsim 10^{8}\;\rm{\sec}\;.
\label{eq:hadron_constraint_strong}
\end{equation}
Another constraint comes from the photo-dissociation of light elements
by the NLSP decay. A detailed investigation on this effect was done in
Ref.~\cite{Kawasaki:2000qr}, and the result is available in Fig.~2 of
Ref.~\cite{Kawasaki:2000qr} : 
\begin{eqnarray}
B_{em}\times m_{\rm NLSP}Y_{\rm NLSP}\lsim 10^{-12}\GEV\quad\mbox{for}\quad
\tau_{\rm NLSP}\gsim 10^{6}{\rm sec}\;,
\label{eq:photon_constraint}
\end{eqnarray}
where $B_{em}$ is the branching ratio into electromagnetic
components\footnote{
For $10^{4}{\rm sec}\lsim \t_{\rm NLSP}\lsim 10^{6}{\rm sec}$, this
constraint is rather weak; 
$B_{em}\times m_{\rm NLSP}Y_{\rm NLSP}\lsim
10^{-(5-10)}$~GeV~\cite{Kawasaki:2000qr}.  
}.

The SUSY standard model (SSM) has various candidates for the NLSP, that
is, wino, Higgsino, squark, gluino, bino and scalar lepton (slepton).
All candidates besides the bino and the scalar lepton have dominant hadronic
decay modes and hence they are subject to the new constraint
Eq.~(\ref{eq:hadron_constraint}) from the BBN. 
Our numerical calculation shows that the yields of those particles are  
\begin{equation}
m_{\rm wino} Y_{\rm wino} \sim 10^{-11.2} \GEV
\bigg(\frac{m_{\rm wino}}{100\GEV}\bigg)^2,
\label{eq:relic1}
\end{equation}
\begin{equation}
m_{\rm Higgsino} Y_{\rm Higssino} \sim 10^{-10.7} \GEV
\bigg(\frac{m_{\rm Higgsino}}{100\GEV}\bigg)^2,
\label{eq:relic2}
\end{equation}
\begin{equation}
m_{\rm squark} Y_{\rm squark}\sim 10^{-11.4} \GEV
\bigg(\frac{m_{\rm squark}}{100\GEV}\bigg)^2,
\label{eq:relic3}
\end{equation}
\begin{equation}
m_{\rm gluino}Y_{\rm gluino} \sim 10^{-12.3} \GEV
\bigg(\frac{m_{\rm gluino}}{100\GEV}\bigg)^2.
\label{eq:relic4}
\end{equation}
For our purpose, we have calculated the relic densities of the candidates
for the NLSP by using {\it micrOMEGAs} computer
code~\cite{Belanger:2001fz}, 
which includes all possible co-annihilation effects.\footnote{ 
The above numerical expressions for the NLSP abundance depend on
parameters of models and may be enhanced by about a factor of 3,
which does not, however, affect the following discussion. 
Here, we have assumed that the annihilation processes of the NLSP's do
not take place near poles of some particles. 
We have  neglected also the nonperturbative QCD effects for the gluino 
annihilation process. 
}
We see that non of them satisfies the cosmological constraint 
Eq.~(\ref{eq:hadron_constraint}) except for the light gluino of mass
$\lsim 70$~GeV.
However, as pointed out in the introduction, this interesting
possibility was already excluded by the present accelerator experiments
\cite{Hagiwara:fs}.\footnote{
OPAL and CDF data exclude the existence of (quasi)stable gluino in the
mass range $3$~GeV $\lsim m_{\rm gluino} \lsim 23$~GeV and $35$~GeV
$\lsim m_{\rm gluino} \lsim 130$~GeV \cite{Hagiwara:fs}. 
Thus, the mass of the stable gluino is still allowed between
23~GeV and 35~GeV. 
}

We should note here that if lifetimes of the NLSP's are shorter than
$10^2$~sec the constraint from the hadronic effects on the BBN becomes
weaker \cite{Kohri} as 
\begin{eqnarray}
B_{h}\times m_{\rm NLSP}Y_{\rm NLSP}\lsim 4\times 10^{-9}\GEV\quad
{\mbox{for}}\quad 10^{-1}\;{\rm sec}\lsim \tau_{\rm NLSP}\lsim 10^{2}\;
\rm{\sec}\;.
\label{eq:hadron_constraint_weak}
\end{eqnarray}
We find that this weaker constraint can be satisfied when the NLSP mass
is smaller than a few TeV (see
Eq.~(\ref{eq:relic1})--(\ref{eq:relic4})).   
However, we see from Eq.~(\ref{eq:thermal_relic})\footnote{    
Here, we set $m_{\rm gluino}\simeq m_{\rm NLSP}$ in
Eq.~(\ref{eq:thermal_relic}) for a conservative estimation. 
}
 and (\ref{eq:NLSP_lifetime}) that $\Omega^{\rm th}_{3/2}h^2$ is always
 larger than $\Omega_{\rm DM}h^2$ for $\tau_{\rm NLSP} \lsim 10^2$~sec
 and  $T_R \gsim 3\times 10^{9}$~GeV.
As a result, no NLSP candidate which have dominant hadronic decay modes
satisfies all the constraints, Eq.~(\ref{eq:hadron_constraint_weak}),
$\Omega^{\rm th}_{3/2}h^2 < \Omega_{\rm DM}h^2$ and $T_R\gsim 3\times
10^{9}$~GeV, simultaneously.

On the other hand, the bino and the scalar lepton are still possible
candidates for the NLSP, since hadronic energy releases from their decays 
are very small~\cite{Feng,Kawasaki:1994bs} and they are not subject to
the strong constraint Eq.~(\ref{eq:hadron_constraint}).
However, the bino NLSP is not interesting, since its electromagnetic
decay violates the constraint Eq.~(\ref{eq:photon_constraint}) as
pointed out in \cite{BBP}.
Therefore, we concentrate ourselves to the case of the scalar lepton
NLSP in the subsequent sections.  
\section{Upper bound on the gluino mass with the scalar charged lepton
 NLSP}  
There are two candidates for the scalar lepton NLSP. 
One is the scalar charged lepton and the other the scalar neutrino.
The scalar neutrino decays into a gravitino and a neutrino.
The effects on BBN from the produced high-energy neutrino will be
discussed in the next section.

We consider, in this section, the consequence of the constraints we have  
discussed in the previous section, taking the lightest scalar charged
lepton (probably stau) to be the NLSP.\footnote{
The possibility that the stau is the NLSP was considered in the framework
of gauge-mediation model~\cite{Asaka:2000zh}.
} 
Constraints for the stau NLSP come from the photo-dissociation of light
elements by the stau decay, and then we determine the upper bound on the
gluino mass $m_{\rm gluino}$ from the relic gravitino abundance in the
following procedure.
(In the slepton NLSP case, the hadronic
contribution dominantly comes from three- and four-body decay channels, 
such as $lZ\widetilde{G}$ and
$lq\bar{q}\widetilde{G}$~\cite{Feng}. 
The branching ratios for these modes are highly suppressed
as $B_{h}=10^{-3}\mbox{--}10^{-5}$. However, even in this case,
if we take the constraint from $^6$Li/$^7$Li very seriously,
the allowed regions that  we will present in the following 
are likely to be reduced.
In the rest of the paper, we take a conservative point of view
assuming this constraint to be avoided.)

\begin{figure}[htb]
 \begin{center}
\begin{minipage}{0.45\linewidth}
\begin{center}(a)
 \includegraphics[width=\linewidth]{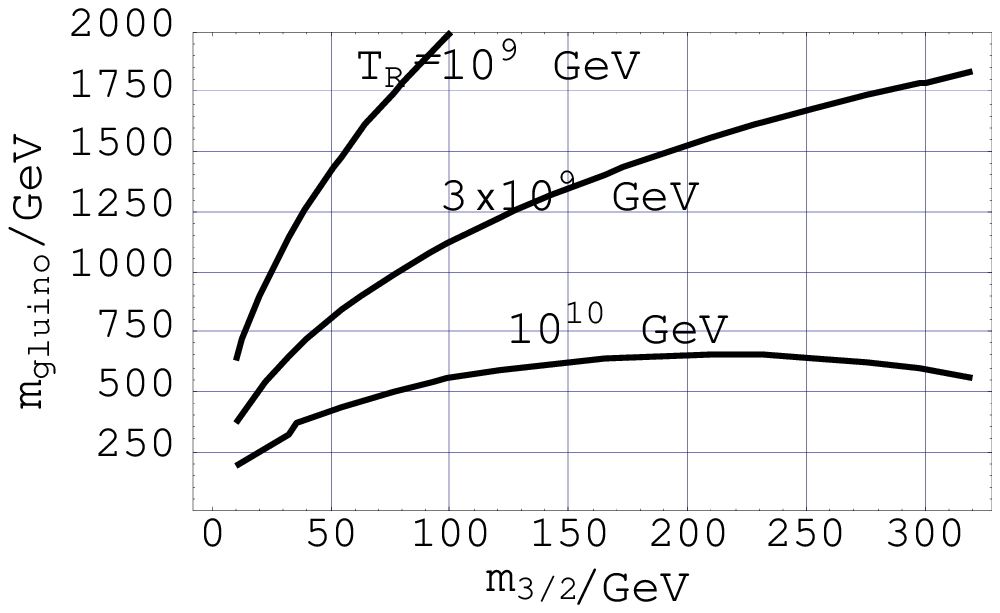}
\end{center}
\end{minipage}
\begin{minipage}{0.45\linewidth}
\begin{center}(b)
  \includegraphics[width=\linewidth]{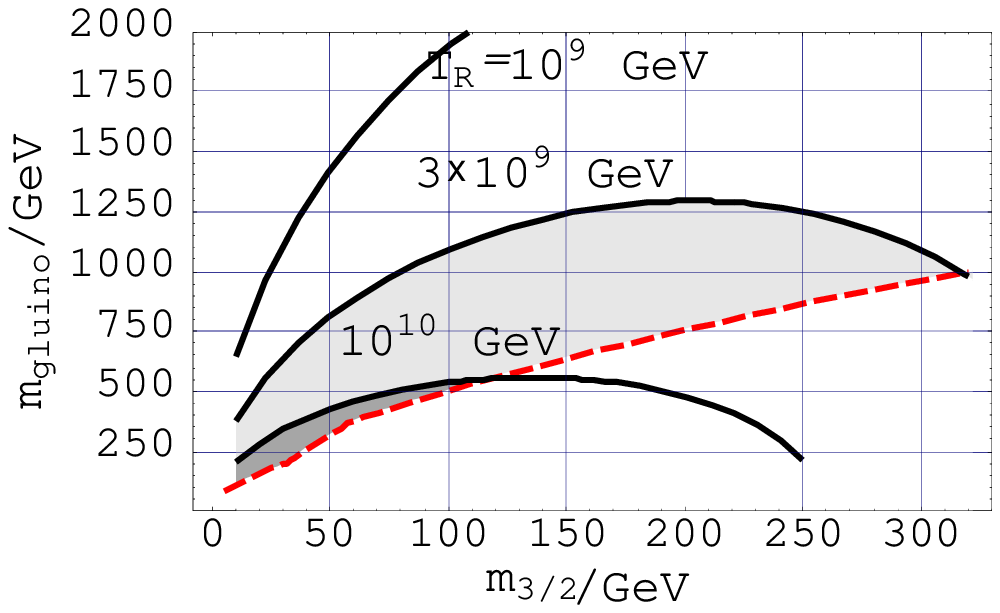}
\end{center}
\end{minipage}
\caption{The upper bound on the gluino mass at a given gravitino mass.
The solid lines show the upper bounds on the gluino mass for the reheating
 temperature $T_R = 10^{10}$~GeV, $3\times 10^9$~GeV and $10^9$~GeV from
 the bottom up, respectively.
The dashed line denotes the lower limit on the stau NLSP mass for a
 given  gravitino mass.
We include nonthermal relic abundance of the gravitino in the panel
  (b) (tan$\beta=30$).   
}
\label{fig:stau}
\end{center}
\end{figure}

Since the relic abundance of the gravitino $\Omega^{\rm th}_{3/2}h^2$
must be smaller than $\Omega_{\rm DM}h^2 \simeq 0.11$, we obtain the
upper bound on the gluino mass at a given gravitino mass from
Eq.~(\ref{eq:thermal_relic}).
We show the resultant upper bound on $m_{\rm gluino}$
in Fig.~\ref{fig:stau}-(a).   
We see that this constraint does not provide a correct upper bound on 
$m_{\rm gluino}$, since in addition to the thermal relic abundance of
the gravitino $\Omega^{\rm th}_{3/2}h^2$, there is a nonthermal
contribution from the late-time decay of the stau NLSP, $\Omega^{\rm
nonT}_{3/2}h^2$. 

To determine $\Omega^{\rm nonT}_{3/2}h^2$ at a given gravitino mass, we
calculate the abundance of the stau NLSP,  $m_{\rm stau}Y_{\rm stau}$,
before its decay at a given stau mass $m_{\rm stau}$.
Our numerical result\footnote{
Here, we have assumed the gravity-mediated supersymmetry breaking at the 
grand unified theory (GUT) scale $\simeq 2\times10^{16}$~GeV. 
We have set masses squared of the sleptons $m_{\rm slepton}^2 +
(200\GEV)^2$ for the first and second generations and $m_{\rm
slepton}^2$ for the third generation at the GUT scale, where $m_{\rm
slepton}$ is a universal slepton mass.  
This boundary condition is favorable to avoid accidental
co-annihilations.  
In this sense our estimation on $Y_{\rm stau}$ is conservative.
}
is $m_{\rm stau}Y_{\rm stau} \sim 10^{-10.3} (m_{\rm stau}/100\GEV)^2$
for tan$\beta = 30$, where tan$\beta$ is a ratio of vacuum expectation
values of the two neutral Higgs bosons $(H_u^0,H_d^0)$, tan$\beta \equiv
\vev{H_u^0}/\vev{H_d^0}$.\footnote{
The above numerical expression for  $m_{\rm stau}Y_{\rm stau}$ may
depend on parameters of models, although the following discussion will
not change very much.
For example, we find $m_{\rm stau}Y_{\rm stau} \sim 10^{-10.1} (m_{\rm
stau}/100\GEV)^2$ for tan$\beta =10$. 
It also may be reduced by a factor of three for the boundary
condition which gives a larger mass to the right hand scalar lepton
than the left one at the GUT scale.    
}
Then, we can determine the upper bound on the lifetime of the stau NLSP
from the Fig.~2 of Ref.~\cite{Kawasaki:2000qr} for a given $m_{\rm
stau}$, which determines the upper bound on the gravitino mass $m_{3/2}$
(see Eq.~(\ref{eq:NLSP_lifetime})).  
By reversing this argument, we can obtain the lower bound on the stau
mass and hence its abundance, $\Omega_{\rm stau}h^2$, at a given
gravitino mass.  
By converting this lower bound on $\Omega_{\rm stau}h^2$ to the
 $\Omega_{3/2}^{\rm nonT}h^2$ by  $\Omega_{3/2}^{\rm nonT}h^2 =
 (m_{3/2}/m_{\rm stau})\Omega_{\rm stau}h^2$, we obtain the
 lower bound on the $\Omega_{3/2}^{\rm nonT}h^2$  at a given gravitino
 mass. 
Finally, we obtain the upper bound on the gluino mass for each set of 
$(m_{3/2},T_{R})$ in order to make the total gravitino relic density
$\Omega_{3/2}h^2=\Omega_{3/2}^{\rm th}h^2+\Omega_{3/2}^{\rm nonT}h^2$
 not to exceed the WMAP result, $\Omega_{\rm DM}h^2\simeq 0.11$.

The result of the above procedure is given in Fig.~\ref{fig:stau}-(b),
which shows the upper bound on the gluino mass for a given gravitino
mass. 
We see that the upper bound reaches\footnote{
If one adopts the reheating temperature $T_R\gsim 10^{10}$~GeV for the
leptogenesis~\cite{BBP}, we find the upper bound on the gluino mass to
be $600$~GeV.   
}
1.3~TeV at $m_{3/2}\simeq 200$~GeV for the reheating temperature $T_R
\gsim 3\times 10^{9}$~GeV and tan$\beta$ =30.\footnote{ 
We also find that the upper bound on $m_{\rm gluino}$ reaches 1.1~TeV at
$m_{3/2}\simeq 160$~GeV for $T_R \gsim 3\times 10^{9}$~GeV and tan$\beta$=10.
}
 
Here, we comment on the falling-off behavior of the upper bound in the
region of $m_{3/2} \gsim 200$~GeV. 
This  behavior comes from the fact that the nonthermal production of the 
gravitino from the stau decay becomes dominant.
Namely, in the region of $m_{3/2}\gsim 200$~GeV the gravitino produced
in the stau decay dominates over the relic gravitino produced just after
the inflation. 
In this sense the mass density of the gravitino is given by the
low-energy parameters and it is almost independent of the reheating
temperature $T_R$.
Thus, we find a gravitino DM scenario becomes more attractive in this 
falling-off region. 

\begin{figure}[htb]
\begin{center}
 \includegraphics[width=0.6\linewidth]{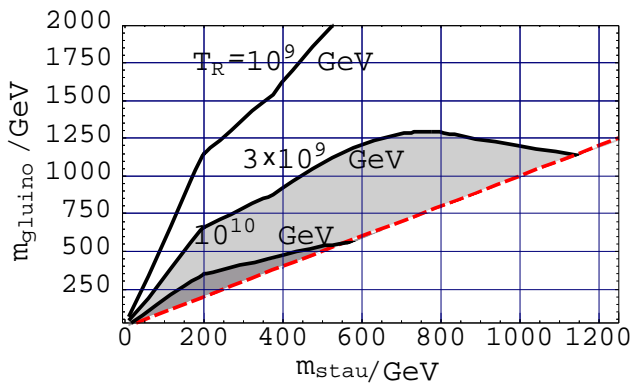}
\end{center}
\vspace{-1cm}
\caption{The upper bound on the gluino mass at a given stau mass
 (tan$\beta=30$). 
The solid lines show the upper bounds on the gluino mass for the reheating
 temperature $T_R = 10^{10}$~GeV, $3\times 10^9$~GeV and $10^9$~GeV from
 the bottom up, respectively.
The dashed line denotes the lower limit on the gluino mass at a given
 stau mass, $m_{\rm gluino} = m_{\rm stau}$.
 }
\label{fig:staugluino}
\end{figure}

Finally, we should note that the shaded regions in the
Fig.~\ref{fig:stau} are conservative ones, thus, they are not always
allowed for a given value of ($m_{\rm stau}$, $m_{\rm gluino}$), since
we use the lower bound on $m_{\rm stau}$ at a given gravitino mass to
estimate $\Omega_{3/2}^{\rm nonT}h^2$. 
Alternatively, we can obtain the upper bound on $m_{\rm gluino}$ for a
given $m_{\rm stau}$ in the following procedure.
As discussed above, we can obtain the $\Omega_{\rm stau}h^2$ and
the upper bound on $m_{3/2}$ at a given $m_{\rm stau}$.
Then, we search the upper bound on $m_{\rm gluino}$ which satisfies
$\Omega^{\rm th}_{3/2}h^2 + \Omega^{\rm nonT}_{3/2}h^2 \lsim \Omega_{\rm
DM}h^2\simeq 0.11$ for each given $m_{\rm stau}$ within the above 
gravitino mass bound.  
We show in Fig.~\ref{fig:staugluino} the allowed parameter region in the 
($m_{\rm stau}$, $m_{\rm gluino}$) plane.

\section{Upper bound on the gluino mass with the scalar neutrino NLSP}   
In this section, we consider the constraints on another candidate for
the NLSP, a scalar neutrino. 
A constraint on the scalar neutrino NLSP comes from the destruction of 
the light elements produced at the BBN epoch, which is caused by the
high energy neutrino injection from the scalar neutrino decay.
The high energy neutrino scatters off the background neutrino and
produce a lepton--antilepton pair, then it produces many soft photons
through electromagnetic cascade processes and destructs the light
elements.   
A detailed analysis on the effects of the high energy neutrino injection
was made in Ref.~\cite{Kawasaki:1994bs} for the case of the scalar
neutrino LSP.
We convert the constraint on the reheating temperature in the Fig.~2 in
\cite{Kawasaki:1994bs} to the NLSP abundance as
\begin{eqnarray}
 m_{\rm NLSP}\times Y_{\rm NLSP}\lsim 10^{-8}\GEV
\quad\mbox{for}\quad
10^3~{\rm sec}\lsim\tau_{\rm NLSP}\lsim 10^9~{\rm sec}.
\label{eq:neutrino}
\end{eqnarray}
Our numerical result for the abundance of the scalar neutrino is
$m_{\tilde{\nu}}Y_{\tilde{\nu}} \sim 10^{-10.8} (m_{\rm \tilde{\nu}} 
/100\GEV)^2$, where subscript $\tilde{\nu}$ denotes the scalar
neutrino. 
We see that the scalar neutrino of mass $\lsim 3$~TeV satisfies the
cosmological constraint Eq.~(\ref{eq:neutrino}).
As we will see in the following discussion, this constraint is not
significant to determine the upper bound on the gluino mass which is
given by the requirement for the total gravitino relic density not to 
exceed the WMAP result, $\Omega_{\rm DM}h^2 \simeq 0.11$.  

To obtain the upper bound on the gluino mass, let us remember that
the lower bound on the scalar neutrino mass is fixed (by the definition 
of the NLSP) as $m_{\tilde{\nu}} > m_{3/2}$ at a given gravitino
mass.\footnote{     
For the scalar neutrino of $m_{\tilde{\nu}}\lsim m_{3/2}$ (the scalar
neutrino LSP), the constraint from the direct detection experiment for
the dark matter is very stringent~\cite{Falk:1994es}. 
}
Then, it determines the lower bound on the relic abundance of the scalar
neutrino, $\Omega_{\tilde{\nu}}h^2$, at a give gravitino mass.
As in the previous section, this lower bound on
$\Omega_{\tilde{\nu}}h^2$ give rise to lower bound of the $\Omega^{\rm 
nonT}h^2$ at a given gravitino mass.  
Finally, we obtain the upper bound on the gluino mass for each set of
($m_{3/2}$, $T_R$) in order to make the total gravitino relic density
$\Omega_{3/2}h^2 = \Omega^{\rm th}_{3/2}h^2 + \Omega^{\rm nonT}_{3/2}
h^2$ not to exceed the WMAP result, $\Omega_{\rm DM}h^2 \simeq 0.11$.

\begin{figure}[htb]
 \begin{center}
\begin{minipage}{0.45\linewidth}
\begin{center}(a)
 \includegraphics[width=\linewidth]{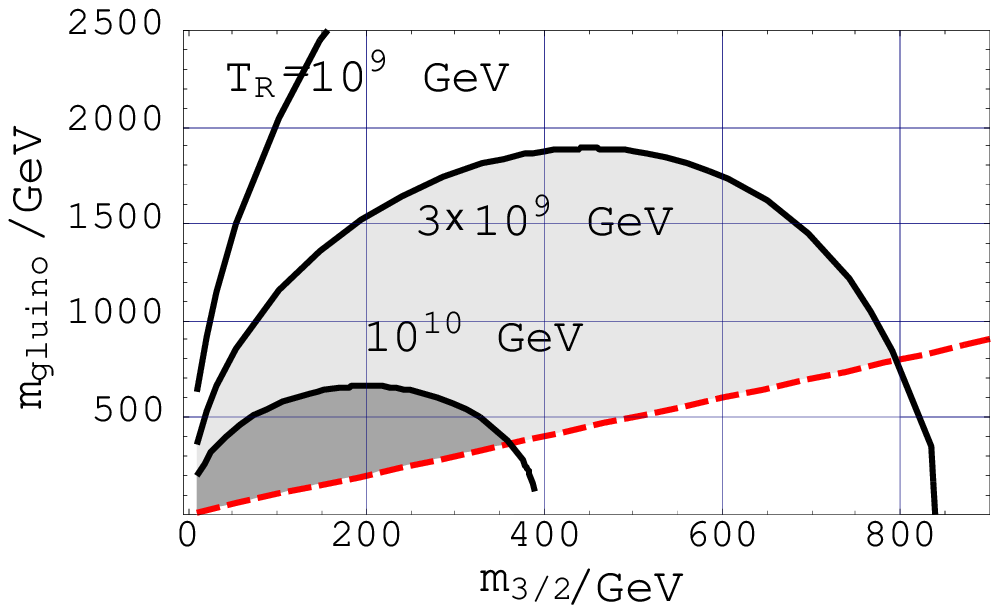}
\end{center}
\end{minipage}
\begin{minipage}{0.45\linewidth}
\begin{center}(b)
  \includegraphics[width=\linewidth]{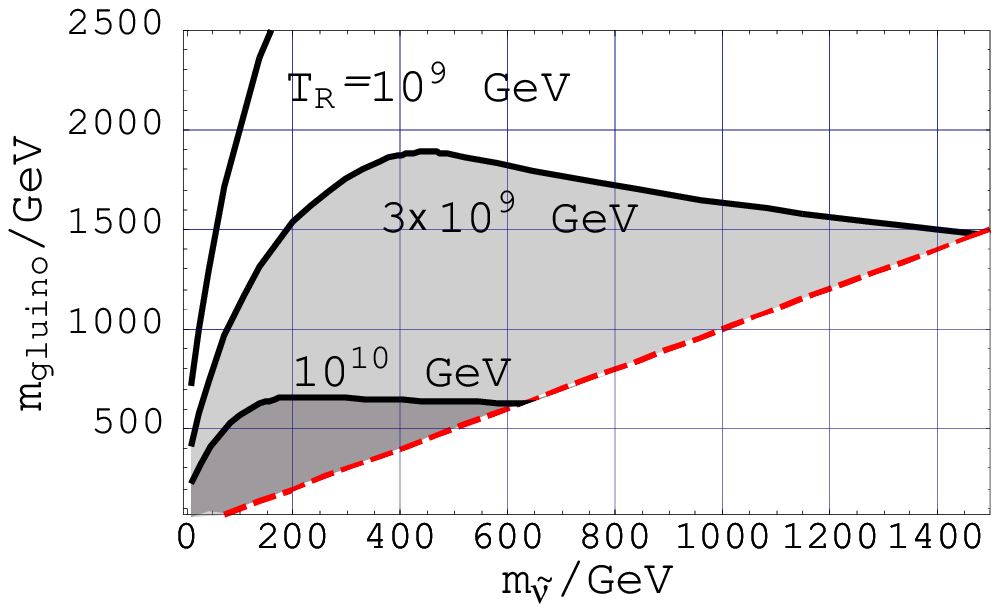}
\end{center}
\end{minipage}
\end{center}
\caption{The upper bound on the gluino mass at a given gravitino mass
(panel (a)) and at a given scalar neutrino mass (panel (b)).
The solid lines show the upper bounds on the gluino mass for the reheating
 temperature $T_R = 10^{10}$~GeV, $3\times 10^9$~GeV and $10^9$~GeV from
 the bottom up, respectively.
The dashed line in the panel (a) denotes the lower limit on the scalar
 neutrino NLSP mass ($m_{\tilde{\n}}=m_{3/2}$), and the one in the panel
 (b) denotes the lower limit on the gluino mass 
($m_{\rm gluino}=m_{\tilde{\nu}}$). 
Here, we use the relic abundance of the scalar neutrino as
 $m_{\tilde{\nu}} Y_{\tilde{\nu}} \sim 10^{-10.8} (m_{\tilde{\nu}}/100\GEV)^2$.} 
\label{fig:sneutrino}
\end{figure}

The result of the above procedure is given in Fig.~\ref{fig:sneutrino}, 
which shows the upper bound on the gluino mass at a given gravitino
mass (panel (a)) and at a given scalar neutrino mass (panel (b)).  
We see that the upper bound reaches 1.8~TeV at $m_{3/2}\simeq 450$~GeV for
the reheating temperature $T_R \gsim 3\times 10^{9}$~GeV. 
  
\section{Conclusions}
 A recent detailed analysis \cite{Kohri} of the hadronic effects of the  
 gravitino decay on the BBN leads to a stringent constraint on the
 reheating temperature of the inflation as $T_R \lsim 10^6$~GeV. 
This constraint contradicts the condition for the thermal leptogenesis,
that is $T_R \gsim 3\times 10^{9}$~GeV. 
A solution to this serious problem is provided \cite{gravitinoLSP} if
the gravitino is stable LSP. 
However, it depends on nature of the NLSP if this solution works or
not. 
We have shown in this letter that the consistent candidate for the NLSP
 is a scalar charged lepton or a scalar neutrino.  
We have found that there are upper bounds on the gluino  mass  $\lsim
 1.3$~TeV and $1.8$~TeV, for the former and the latter case,\footnote{
We see from Fig.~\ref{fig:staugluino} and Fig.~\ref{fig:sneutrino}, that
 we can not require the universality for three gaugino masses at the GUT
 scale for a large part of the allowed region, since the bino must be 
 heavier than the scalar lepton.  
}
 respectively, so that the density of the  gravitino do not exceed the
 observed dark matter density $\Omega_{\rm  DM}h^2\simeq 0.11$ for
 $T_R\gsim 3\times 10^{9}$~GeV.

In the present analysis we have used the perturbative calculation in
evaluating the yield of relic gluino.
However, it should be kept in mind that the nonperturbative QCD dynamics
may increase the annihilation cross section of gluino which decreases
the yield of gluino NLSP~\cite{Baer:1998pg}.
For instance, if it increases the annihilation cross section by a factor
of 100, the gluino of mass $\lsim 300$~GeV satisfies the BBN constraint
Eq.~(4).
In this case the reheating temperature $T_R$ can be easily taken above
$T_R\simeq 3\times 10^{9}$~GeV.

\section*{Acknowledgment}

The authors wish to thank M.Kawasaki and K.Kohri for a useful
discussion. 
M.F. would like to thank the Japan Society for the Promotion of Science
for financial support.
This work is partially supported by Grand-in-Aid Scientific Research (s)
14102004.

\end{document}